\documentclass[showpacs,twocolumn,aps,prl,superscriptaddress]{revtex4}
\usepackage{mathrsfs}
\usepackage{amsmath}
\usepackage{amsfonts}
\usepackage{amssymb}
\usepackage{amsthm}
\usepackage{graphicx}

\begin{document}

\title{Optimal time-dependent polarized current pattern for fast domain wall
propagation in nanowires: Exact solutions for biaxial and uniaxial
anisotropies}
\author{P. Yan}
\affiliation{Physics Department, The Hong Kong University of Science and Technology,
Clear Water Bay, Hong Kong SAR, China}
\author{Z. Z. Sun}
\affiliation{Institute for Theoretical Physics, University of Regensburg, D-93040
Regensburg, Germany}
\author{J. Schliemann}
\affiliation{Institute for Theoretical Physics, University of Regensburg, D-93040
Regensburg, Germany}
\author{X. R. Wang}
\email[Corresponding author: ]{phxwan@ust.hk}
\affiliation{Physics Department, The Hong Kong University of Science and Technology,
Clear Water Bay, Hong Kong SAR, China}

\begin{abstract}
One of the important issues in nanomagnetism is to lower the current
needed for a technologically useful domain wall (DW) propagation
speed. Based on the modified Landau-Lifshitz-Gilbert (LLG) equation
with both Slonczewski spin-transfer torque and the field-like
torque, we derive the optimal spin current pattern for fast DW
propagation along nanowires. Under such conditions, the DW velocity
in biaxial wires can be enhanced as much as ten times compared to
the velocities achieved in experiments so far. Moreover, the fast
variation of spin polarization can help DW depinning. Possible
experimental realizations are discussed.
\end{abstract}

\maketitle

%\date{\today}

%\email[To whom correspondence should be addressed. Electronic
%address: ] {phxwan@ust.hk}

%\keywords{Domain-wall motion, magnetic nanowires,spintronics}
%\date{\today}
% 75.60.Jk Magnetization reversal mechanisms\\
%75.75.+a Magnetic properties of nanostructures \\
%75.60.Ch Domain walls and domain structure
%85.75.-d Magnetoelectronics; spintronics:
%devices exploiting spin polarized transport or integrated magnetic fields
%85.70.Ay Magnetic device characterization, design and modeling \\

%----------------------------------------------------------------%

\emph{Introduction}$\emph{.}{\LARGE -}$Fast magnetic domain wall (DW)
propagation along nanowires by means of electrical currents is presently
under intensive study in nano-magnetism, both experimentally \cite%
{Parkin,Yamaguchi,Klaui1,Erskine1,Hayashi} and theoretically \cite%
{Tatara1,Barnes,Brataas}. In addition to the technological interest such as
race track memory \cite{Parkin}, DW dynamics is also an interesting
fundamental problem. The dynamics of a single DW can be qualitatively
understood from one-dimensional (1D) analytical models \cite%
{Walker,Slonczewski,Thiaville} that predict a rigid-body propagation
below the Walker breakdown and an oscillatory motion above it
\cite{Walker,xrw1}. The latter process is connected with a series of
complicated cyclic transformations of DW structure and a drastic
reduction of the average DW velocity. The Walker limit is thus the
maximum velocity at which DW can propagate in magnetic nanowires
without changing its inner structure. From a technological point of
view, such a limit seems to represent a major obstacle since the
fidelity of data transmission may depend on preserving the DW
structure while the utility requires speeding up the DW velocity
adequately. Various efforts have been made to overcome this limit
through geometry design. For instance, Lewis \emph{et al}.
\cite{Lewis} proposed a \textquotedblleft chirality
filter\textquotedblright\ consisting of a cross-shaped trap to
preserve the DW structure. Yan \emph{et al}. \cite{Yan} demonstrated
the removal of Walker limit via a micromagnetic study on the
current-induced DW motion in cylindrical Permalloy nanowires. In
this Letter we investigate other ways to substantially increase the
DW velocity avoiding the Walker breakdown.

A DW propagates under a spin-polarized current through angular momentum
transfer from conduction electrons to the local magnetization, known as the
spin transfer torque (STT) \cite{Slon}, which is different from magnetic
field driven DW propagation originated from the energy dissipation \cite%
{xrw1,Sun}. Generally, two types of spin
torques are considered:
the Slonczewski torque \cite{Slon} ($a-$term) $\mathbf{T}_{\text{a}%
}=-\gamma \frac{a_{J}}{M_{s}}\mathbf{M}\times \left( \mathbf{M}\times
\mathbf{s}\right) $ and the field-like torque \cite{Heide,szhang2} ($b-$%
term) $\mathbf{T}_{\text{b}}=-\gamma b_{J}\mathbf{M}\times \mathbf{s}$,
where $\gamma =\left\vert e\right\vert /m_{e}$, $\mathbf{M}$, $%
M_{s}=\left\vert \mathbf{M}\right\vert $, and $\mathbf{s}$ are the the
gyromagnetic ratio, magnetization of the magnet, the saturation
magnetization, and the spin polarization direction of itinerant electrons,
respectively. $a_{J}$ and $b_{J}$ depend on current density $j_{e}$ and spin
polarization $P$. Theory predicts \cite{Slon,szhang2} that $a_{J}=Pj_{e}\hbar
/2d\left\vert e\right\vert M_{s}$ and $b_{J}=\beta a_{J}$, where $d$ is the
thickness of the free magnetic layer. $\beta $ is a small dimensionless
parameter that describes the relative strength of the field-like torque to
the Slonczewski torque. The value of $\beta $ is sensitive to the thickness
of the free layer and the decay length of the transverse component of the
spin accumulation inside the free layer as discussed in Ref. \cite{szhang2}.
The typical value of $\beta $ ranges from $0$ to $0.5$ \cite{szhang2,Stiles}%
. In the conventional case of current along the nanowire with biaxial magnetic
anisotropy, the $a-$term is
incapable of generating a sustained DW motion, except for a very large
current, while the $b-$term can drive a DW to propagate \cite{Tatara1}.
Unfortunately, the $b-$term is usually much smaller than $a-$term \cite%
{Erskine1,Hayashi}. A large current density is needed in order to reach a
technologically useful DW propagation velocity \cite{Parkin}, but the
associated Joule heating and DW structure collapse could affect device
performance. We show that the problem can be solved if one uses an optimal
polarized current pattern.

In this Letter, our focus is on the optimal spin-polarized electric current
pattern for fast DW propagation along nanowires. For usual magnetic
materials, our theoretical results show that the DW velocity can be enhanced
by as large as ten times in comparison with DW velocity driven by the
conventional constant current in existing experiments. Moreover, the
ultrafast change of spin polarization can be used to de-pin a DW.

\emph{Model}$\emph{.}{\LARGE -}$ The internal magnetic energy of a nanowire
can be formulated as
\begin{equation}
U\left[ \mathbf{M}\right] =\int d^{3}x\Bigl (\frac{J}{2}\left[ \left( \nabla
\theta \right) ^{2}+\sin ^{2}\theta \left( \nabla \phi \right) ^{2}\right]
+w\left( \theta ,\phi \right) \Bigr ),  \label{Free energy}
\end{equation}%
where $\theta $ and $\phi $ are the polar angle and azimuthal angle of the
local magnetization $\mathbf{m=M}/M_{s},$ respectively. $J$ and $w$ are the
exchange energy constant and energy density due to all kinds of
anisotropies, respectively. The dynamics of $\mathbf{M}$ is governed by the
LLG equation \cite{Gilbert}:
\begin{equation}
\frac{\partial \mathbf{M}}{\partial t}=-\gamma \mathbf{M}\times \mathbf{H}%
_{eff}+\frac{\alpha }{M_{s}}\mathbf{M}\times \frac{\partial \mathbf{M}}{%
\partial t}+\mathbf{T}_{\text{STT}},  \label{LLG}
\end{equation}%
here $\mathbf{H}_{eff}=-\frac{1}{\mu _{0}}\delta U/\delta \mathbf{M}$ is the
effective magnetic field and $\alpha $ is the phenomenological Gilbert
damping constant \cite{Gilbert}. $\mathbf{T}_{\text{STT}}$ is the STT with
both the Slonczewski-type and the field-like terms.

\emph{Biaxial anisotropy}$\emph{.}{\LARGE -}$Considering a biaxial
anisotropy $w\left( \theta ,\phi \right) =-\frac{K}{2}m_{z}^{2}+\frac{%
K_{\perp }}{2}m_{x}^{2}$ with the easy axis along $\hat{z}$ direction and
the hard axis along $\hat{x}$ direction, the effective field takes a form of
$\mathbf{H}_{eff}=\frac{1}{\mu _{0}M_{s}}\left( Km_{z}\hat{z}-K_{\perp }m_{x}%
\hat{x}\right) +\frac{J}{\mu _{0}M_{s}^{2}}\frac{\partial ^{2}\mathbf{M}}{%
\partial z^{2}}.$ Here $K$ and $K_{\perp }$ describe energetic anisotropies
along easy $\hat{z}$ axis and hard $\hat{x}$ axis, respectively. We assume
that all local spins lie in a fixed plane called DW plane, i.e., $\phi
\left( z,t\right) =\phi \left( t\right) ,$ which should be checked
self-consistently late. In the spherical coordinates, Eq. \eqref{LLG}
becomes
\begin{eqnarray}
\dot{\theta}+\alpha \sin \theta \dot{\phi} &=&\gamma a_{J}\left( s_{\theta
}+\beta s_{\phi }\right) +\frac{\gamma K_{\perp }}{2\mu _{0}M_{s}}\sin
\theta \sin 2\phi ,  \notag \\
&&  \label{LLG0} \\
\sin \theta \dot{\phi}-\alpha \dot{\theta} &=&\gamma a_{J}\left( s_{\phi
}-\beta s_{\theta }\right)   \notag \\
&&-\gamma \frac{J\frac{\partial ^{2}\theta }{\partial z^{2}}-\frac{\sin
2\theta }{2}\left( K+K_{\perp }\cos ^{2}\phi \right) }{\mu _{0}M_{s}},
\label{LLG1}
\end{eqnarray}%
where $s_{r},$ $s_{\theta },$ and $s_{\phi }$ are three components of unit
spin vector $\mathbf{s}$ in spherical coordinates. The DW profile satisfies $%
J\frac{\partial ^{2}\theta }{\partial z^{2}}-\frac{\sin 2\theta }{2}\left(
K+K_{\perp }\cos ^{2}\phi \right) =0$ with boundary condition of $\theta =0$
and $\pi $ at distance. One obtains the famous Walker's DW motion profile $%
\tan \frac{\theta }{2}=\exp \left( \frac{z-X\left( t\right) }{\Delta }%
\right) ,$ in which $X\left( t\right) $ is the position of the DW center and
$\Delta =\sqrt{\frac{J}{K+K_{\perp }\cos ^{2}\phi }}$ is DW width resulting
from the balance of anisotropy energy and exchange energy \cite{Walker}.
These assumptions are valid under sufficiently low current density which
will be demonstrated later. Substituting this DW profile into Eqs. %
\eqref{LLG0} and \eqref{LLG1}, we have
\begin{eqnarray}
-\frac{\dot{X}}{\Delta }+\alpha \dot{\phi} &=&\gamma \frac{a_{J}\left(
s_{\theta }+\beta s_{\phi }\right) }{\sin \theta }+\frac{\gamma K_{\perp }}{%
2\mu _{0}M_{s}}\sin 2\phi , \\
\alpha \frac{\dot{X}}{\Delta }+\dot{\phi} &=&\gamma \frac{a_{J}\left(
s_{\phi }-\beta s_{\theta }\right) }{\sin \theta }.  \label{motion}
\end{eqnarray}

For a given DW motion, term $s_{\phi }-\beta s_{\theta }$ should be as large
as possible in order to lower the needed current density. Meanwhile,
considering identity $\left( s_{\phi }-\beta s_{\theta }\right) ^{2}+$ $%
\left( s_{\theta }+\beta s_{\phi }\right) ^{2}=\left( 1+\beta ^{2}\right)
\left( 1-s_{r}^{2}\right) ,$ we choose%
\begin{equation}
s_{r}=0,\text{ }s_{\theta }=\cos \eta ,\text{ }s_{\phi }=\sin \eta ,
\label{Pattern1}
\end{equation}%
with optimization parameter $\eta $.

To ensure the spatial-independence of $\dot{X}$ and $\phi ,$ the above
equations require $a_{J}$ to be proportional to $\sin \theta ,$ so we let $%
a_{J}=A_{J}\sin \theta =A_{J}\sec $h$\left( \frac{z-X\left( t\right) }{%
\Delta }\right) $ with a constant $A_{J}.$ Thus we have%
\begin{eqnarray}
\dot{X} &=&\gamma \Delta \frac{\alpha a_{J}^{\prime }-b_{J}^{\prime }}{%
1+\alpha ^{2}}-\Delta \frac{\gamma K_{\perp }}{2\mu _{0}M_{s}\left( 1+\alpha
^{2}\right) }\sin 2\phi ,  \label{Propagation} \\
\dot{\phi} &=&\gamma \frac{a_{J}^{\prime }+\alpha b_{J}^{\prime }}{1+\alpha
^{2}}+\frac{\alpha \gamma K_{\perp }}{2\mu _{0}M_{s}\left( 1+\alpha
^{2}\right) }\sin 2\phi ,  \label{Precession}
\end{eqnarray}%
which describe the DW propagation and the DW plane precession. Here $%
a_{J}^{\prime }\left( \eta \right) =A_{J}\left( \sin \eta -\beta \cos \eta
\right) $ and $b_{J}^{\prime }\left( \eta \right) =A_{J}\left( \cos \eta
+\beta \sin \eta \right) $. The DW width is time-independent when the DW
undergoes a rigid-body propagation with $\phi \left( t\right) =\phi _{0}=$%
constant. In general, DW width $\Delta $ depends on the time through
time-dependence of $\phi .$ The exact rigid-body solutions constitute%
\begin{eqnarray}
\frac{\alpha K_{\perp }}{2\mu _{0}M_{s}}\sin 2\phi _{0}\left( \eta \right)
&=&-\left( a_{J}^{\prime }+\alpha b_{J}^{\prime }\right) ,  \label{Angle} \\
\Delta \left( \eta \right) &=&\sqrt{\frac{J}{K+K_{\perp }\cos ^{2}\phi
_{0}\left( \eta \right) }},  \label{Width} \\
\dot{X}\left( \eta \right) &=&\gamma \frac{a_{J}^{\prime }}{\alpha }\Delta
\left( \eta \right) .  \label{Optimal Velocity}
\end{eqnarray}

The spin current pattern is then described by $\eta .$ Different value leads
to different canted angle, DW width and propagation velocity. It is
straightforward to show that the assumption of rigid-body motion always
holds under condition $A_{J}\sqrt{\left( 1+\alpha ^{2}\right) \left( 1+\beta
^{2}\right) }\leq \frac{\alpha K_{\perp }}{2\mu _{0}M_{s}}.$

Before finding the optimized spin current pattern for maximal velocity,
let's first consider two special cases. The conventional case in the
existing experiments \cite{Fert,Boone} is the constant current density with
electron spin polarization along $z-$axis, i.e., $s_{\theta }=-\sin \theta $
and $s_{\phi }=0,$ which gives the velocity $u_{1}=\gamma \frac{\beta A_{J}}{%
\alpha }\Delta \left( \pi \right) .$ It again shows that Slonczewski torque
is incapable of generating sustained DW propagation while the field-like
torque can. But the velocity is rather small since $\beta \ll 1$ in usual
materials. However, the DW velocity can be greatly enhanced if $a-$term is
involved. This is the case of $\eta =\frac{\pi }{2}.$ It gives the velocity $%
u_{2}=\gamma \frac{A_{J}}{\alpha }\Delta \left( \frac{\pi }{2}\right) .$ In
typical materials \cite{Stiles}, $\beta \sim 0.1,$ so the velocity is 10
times lager than $u_{1}.$ One can see that DW propagation velocity is
greatly enhanced under a modification of the spin polarization and locally
minimized current density pattern.

The maximal velocity $\dot{X}_{\max }=\dot{X}\left( \eta ^{\ast }\right) $
at the optimal parameter $\eta ^{\ast }$ can be found through exact
numerical calculations although a closed analytic form is difficult to
obtain due to the complexity of Eqs. \eqref{Angle}, \eqref{Width}, and %
\eqref{Optimal Velocity}. Factor $\lambda =\dot{X}_{\max }/u_{1}=\frac{\sin
\eta ^{\ast }-\beta \cos \eta ^{\ast }}{\beta }\frac{\Delta \left( \eta
^{\ast }\right) }{\Delta \left( \pi \right) }$ measures the velocity
enhancement. Fig. 1a is the $\beta ^{-1}$ dependence of $\lambda $ for
various damping coefficients and typical magnetic parameters. It is
approximately linear, and insensitive to damping parameter $\alpha $. Fig.
1b is the plot of spatial distribution of $s_{x},$ $s_{y},$ $s_{z}$ and $%
a_{J}$ for the optimized spin current pattern around DW center. We note that
$s_{x},$ $s_{y},$ and $s_{z}$ vary only near the DW center, and reach fixed
values away from the DW. A large perpendicular component $s_{y}$ is required
to achieve large DW\ velocity. The reason is that perpendicular spin
component induces a large effective field $\mathbf{H}_{a}=\frac{a_{J}}{M_{s}}%
\mathbf{M}\times \mathbf{s.}$ Thus the DW moves under the Slonczewski torque
with a large component along wire axis. This finding is consistent with
recent micromagnetic simulations \cite{Khvalkovskiy} showing that the DW
velocity can be greatly increased by applying perpendicular spin
polarizations. It is also very interesting that locally minimized current
density $a_{J}$ is finite only near DW center while it becomes zero at
distance, which should greatly lower the energy consumption.

\begin{figure}[tbph]
\begin{center}
\includegraphics[width=8.cm, height=5cm]{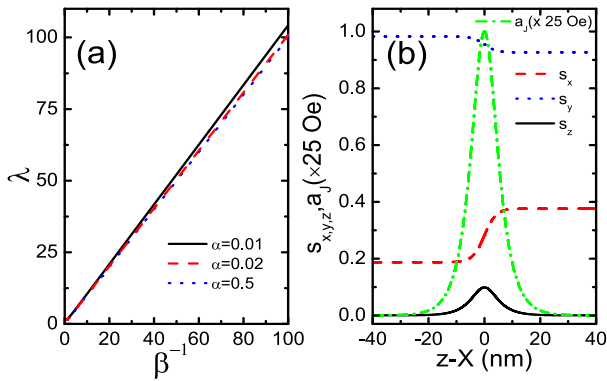}
\end{center}
\caption{(Color online) (a) DW velocity enhancement factor $\protect\lambda$
versus $\protect\beta ^{-1}$ at different damping coefficients $\protect%
\alpha =0.01, 0.02$ and $0.5$. (b) The spatial distribution of $x,y,z$
components of the optimal spin polarization pattern and current density
pattern for maximal DW velocity under $\protect\alpha =0.01$ and $\protect%
\beta =0.1$. The other parameters are using the materials parameters of
Permalloy: $M_{s}=8.6\times 10^{5}$ A/m, $J=1.3\times 10^{-11}$ J/m, $K=500$
J/m$^{3}$, $K_{\perp }=8\times 10^{5}$ J/m$^{3}$ \protect\cite{Parkin}, and
a reasonable value $A_{J}=25$ Oe according to Ref. \protect\cite%
{Khvalkovskiy}.}
\end{figure}

\emph{Depinning}$\emph{.}{\LARGE -}$Besides the advantage of markedly
speeding up the DW velocity, our time-dependent spin current pattern also
implies a possible way to improve the efficiency of DW motion against the
pinning effect. The argument is attributed to an additional force on the
wall due to a fast changing spin direction. It is convenient for us to treat
DW as a quasiparticle \cite{Doring} with mass $m_{\text{w}}=\frac{2S\mu
_{0}^{2}M_{s}^{2}}{\Delta \gamma ^{2}K_{\perp }}$ \cite{Kruger,Bocklage}
when we deal with the effect of pinning. Here $S$\ is the cross section of
the wire. The pinning force $F_{\text{pin}}=-\frac{dE}{dX}$ is expressed by
the pinning potential $E$ and the position $X$ of the wall. Thus for small $%
\phi ,$ Eqs. \eqref{Propagation} and \eqref{Precession} can be simply
decoupled and result in%
\begin{eqnarray}
\frac{F}{m_{\text{w}}} &=&\ddot{X}=\frac{\alpha \gamma K_{\perp }}{\left(
1+\alpha ^{2}\right) \mu _{0}M_{s}}\dot{X}-\frac{1}{m_{\text{w}}}\frac{dE}{dX%
}  \notag \\
&&-\Delta \frac{\gamma ^{2}K_{\perp }}{\mu _{0}M_{s}}\frac{a_{J}^{\prime }}{%
1+\alpha ^{2}}+\gamma \Delta \frac{a_{J}^{\prime }+\alpha b_{J}^{\prime }}{%
1+\alpha ^{2}}\dot{\eta},  \label{Force}
\end{eqnarray}%
where the temporal variation of DW width is neglected. The contributions of
current-induced acceleration are the last two terms. In usual setup ($%
\mathbf{s}$ along the wire axis), the depinning acceleration due to STT is $%
\ddot{X}_{\text{sp}}=-\Delta \frac{\gamma ^{2}K_{\perp }}{\mu _{0}M_{s}}%
\frac{\beta A_{J}}{1+\alpha ^{2}}$ \cite{Bocklage}. Thus, one can
observe that our optimal spin current pattern provides a higher
depinning
acceleration by $10$ times for the current density part since parameter $%
\beta $ is typically around 0.1 \cite{Stiles}. The force $F$ on the
wall does not only depend on the current density but also on the
time derivative of spin direction. The switching-time dependence of
current-induced depinning follows the last term in Eq.
\eqref{Force}. For a fast changing current, i.e., spin variation
rate $\tau \ll \frac{\mu _{0}M_{s}}{\gamma K_{\perp }}\sim 10^{-12}$
s$,$ the contribution of the time derivative term will be
significant. The novelty and importance of our proposal are embodied
in the nature of ultrafast switching-time of spin degree of freedom.
It makes a main difference from the DW depinning due to current
density's rise time \cite{Bocklage} which is limited by the
intrinsic response time of circuit \cite{Heyne}.

\emph{Uniaxial anisotropy}$\emph{.}{\LARGE -}$Let $K_{\perp }=0,$ Eqs. %
\eqref{Propagation} and \eqref{Precession} result in%
\begin{equation}
\dot{X}^{2}+\dot{\phi}^{2}\Delta ^{2}=\frac{\gamma ^{2}\Delta ^{2}A_{J}^{2}}{%
1+\alpha ^{2}}\left( 1+\beta ^{2}\right) ,
\end{equation}%
with DW width parameter $\Delta =\sqrt{J/K}.$

Thus, the largest possible DW propagation velocity is $\dot{X}_{\max
}=\gamma A_{J}\sqrt{\left( 1+\beta ^{2}\right) /\left( 1+\alpha ^{2}\right) }%
\Delta .$ The conventional polarization \cite{Fert,Boone} gives the DW
velocity $u=\gamma A_{J}\frac{1+\alpha \beta }{1+\alpha ^{2}}\Delta .$ As a
result, the DW velocity is enhanced under the optimal spin current pattern
by a factor of $\dot{X}_{\max }/u=$ $\sqrt{1+\left( \frac{\alpha -\beta }{%
1+\alpha \beta }\right) ^{2}}$. We note that the enhancement is not so large
in the uniaxial wire since both $\alpha $ and $\beta $ are far less than $1$
in usual magnetic materials \cite{xrw2}. The physical reason lies in that $a-
$term is capable of generating a sustained DW motion in uniaxial wire, which
is different from biaxial case.

\emph{Discussion}$\emph{.}{\LARGE -}$Although the optimal spin current
pattern for maximum DW velocity is found, it is still an experimental
challenge to generate a temporally and spatially varying spin polarized
current. Interestingly enough, a very recent experiment used spin-polarized
current perpendicular to a nanowire to manipulate DW motion \cite{Boone}.
There are now at least two types of current patterns realizable. The hope is
that our capable experimentalists can one day generate any designed current
pattern. Indeed, there are many theoretical proposals for generating a
designed current pattern. Tao \emph{et al}. \cite{Tao} and Delgado \emph{et
al}. \cite{Delgado} have proposed to use magnetic scanning tunneling
microscopic (STM) tip above a magnetic nanowire to produce localized
spin-polarized current. Experimentally, the control of spin-polarized
current in a STM by single-atom transfer was demonstrated very recently by
Ziegler \emph{et al} \cite{Ziegler}. In summary, our proposed optimal spin
current patterns are difficult to generate now, but their existence does not
violate any fundamental laws and principles. Our results and calculations
will be relevant to experiments when the generation of an arbitrary
spin-polarized current pattern becomes true.

In the above discussions, the spin pumping effect on the DW motion is
neglected because the DW-motion induced current is zero in biaxial wire
since there is no DW plane precession \cite{xrw3} below Walker breakdown and
it is much smaller than the applied external spin-polarized current in
uniaxial wire. According to Ref. \cite{Duine}, the maximum DW-motion
generated electric current density in uniaxial wire is $\left\langle
j_{z}\right\rangle =\frac{\hslash }{eL}\left( \sigma _{\uparrow }-\sigma
_{\downarrow }\right) \frac{\dot{X}_{\max }}{\Delta },$ where $L$ is the
length of the nanowire, $\sigma _{\uparrow }$ and $\sigma _{\downarrow }$
denote the conductivities of the majority and minority electrons. In the
experiments of Beach \emph{et al}. \cite{Erskine1}, $L\sim 40$ $\mu $m, $%
\Delta \sim 20$ nm, and DW velocity $\dot{X}_{\max }\sim 40-100$ m/s. For a
typical conductivity $\sigma _{\uparrow }\sim 10^{6}$ $\Omega ^{-1}$m$^{-1},$
one can find the pumped electric current density no more than $\left\langle
j_{z}\right\rangle \sim 10^{5}-10^{6}$A/m$^{2},$ which is much smaller than
the typically applied current of the order of $10^{10}-10^{12}$A/m$^{2}$ in
experiments \cite{Erskine1}.

\emph{Conclusion}$\emph{.}{\LARGE -}$We propose an optimal spin
current pattern for high DW propagation velocity in magnetic nanowires.
In uniaxial wires this enhancement is of modest size, while in biaxial wires
a factor of a few tens can be achieved.
The nature of ultrafast switching-time of
spin degree of freedom proves to be a novel way to improve the efficiency of
DW motion against the pinning. We expect our proposal will stimulate and
also possibly guide future experiments.

We thank Dr. X.J. Xia and Mr. W. Zhu for valuable discussions. This work is
supported by Hong Kong RGC grants (\#603007, 603508, 604109 and
HKU10/CRF/08- HKUST17/CRF/08), and by  Deutsche Forschungsgemeinschaft
via SFB 689. Z.Z.S. thanks the Alexander von Humboldt
Foundation (Germany) for a grant.

\end{document}